\documentclass[iop]{emulateapj}
\usepackage{graphics}
\usepackage{epstopdf}
\usepackage{amsmath}

\usepackage{graphicx}
\graphicspath{%
    {converted_graphics/}
    {/}
}

\def\gtorder{\mathrel{\raise.3ex\hbox{$>$}\mkern-14mu
     \lower0.6ex\hbox{$\sim$}}}
\def\ltorder{\mathrel{\raise.3ex\hbox{$<$}\mkern-14mu
     \lower0.6ex\hbox{$\sim$}}}

\begin{document}

\title{Accretion disk dynamo as the trigger for X-ray binary state transitions}

\author{Mitchell C. Begelman\altaffilmark{1,2}, Philip J. Armitage\altaffilmark{1,2} and Christopher S. Reynolds\altaffilmark{3,4}}
\altaffiltext{1}{JILA, University of Colorado and NIST, 440 UCB, Boulder, CO 80309-0440, USA}
\altaffiltext{2}{Department of Astrophysical and Planetary Sciences, University of Colorado, Boulder, CO 80309-0391, USA}
\altaffiltext{3}{Department of Astronomy, University of Maryland, College Park, MD 20742-2421, USA}
\altaffiltext{4}{Joint Space-Science Institute (JSI), College Park, MD 20742-2421, USA}

\email{mitch@jila.colorado.edu}
 
\begin{abstract}
Magnetohydrodynamic accretion disk simulations suggest that much of the energy liberated by the magnetorotational instability (MRI) can be channeled into large-scale toroidal magnetic fields through dynamo action. Under certain conditions, this field can dominate over gas and radiation pressure in providing vertical support against gravity, even close to the midplane. Using a simple model for the creation of this field, its buoyant rise, and its coupling to the gas, we show how disks could be driven into this magnetically dominated state and deduce the resulting vertical pressure and density profiles. Applying an established criterion for MRI to operate in the presence of a toroidal field, we show that magnetically supported disks can have two distinct MRI-active regions, separated by a ``dead zone" where local MRI is suppressed, but where magnetic energy continues to flow upward from the dynamo region below. We suggest that the relative strengths of the MRI zones, and the local poloidal flux, determine the spectral states of X-ray binaries. Specifically, ``intermediate" and ``hard" accretion states occur when MRI is triggered in the hot, upper zone of the corona, while disks in ``soft" states do not develop the upper MRI zone. We discuss the conditions under which various transitions should take place and speculate on the relationship of dynamo activity to the various types of quasi-periodic oscillations that sometimes appear in the hard spectral components. The model also explains why luminous accretion disks in the ``soft" state show no signs of the thermal/viscous instability predicted by standard $\alpha-$models.
 
\end{abstract} 


\section{Introduction} 

Despite many successes, standard accretion disk models fail to account for key features of X-ray binary (XRB) phenomenology.  Both analytic $\alpha-$viscosity models \citep{shakura73} and numerical simulations capturing angular momentum transport by magnetorotational instability \cite[e.g.][]{balbus98} predict strong thermal \citep{jiang14b} and viscous instability when radiation pressure dominates other forms of pressure in the disk, yet no evidence of such instability is obviously seen during XRB outbursts.  Standard models also fail to provide a satisfactory explanation for why accretion flows undergo spectral state transitions, where the spectrum switches from soft and thermal to hard and nonthermal, and vice-versa.  Models for the transition typically rely on the existence of two possible disk structures at the same accretion rate \citep{chen95,esin97} --- one cool, dense and thermal and the other hot, tenuous and out of equilibrium --- but provide no compelling mechanism to explain why one is chosen over the other when conditions are such that either could exist.  This problem is most acute at soft-to-hard transitions, which occur at relatively low luminosities where thin disks are expected to be relatively stable. One suggestion to explain this transition posits evaporation of the disk by a hot corona \citep{meyer94,meyer00,rozanska00}, but begs the question of what keeps the corona heated at such a low density that electrons and ions are thermally decoupled, which appears to be necessary in order to prevent the electrons from radiating away most of the heat.  Another difficulty is that the states are not mutually exclusive, but often seem to co-exist in relatively long-lived, hybrid ``intermediate" states.  

Non-standard disk models that include magnetically dominated regions provide a possible solution to these problems. At a minimum the buoyant rise of magnetic field generated within the disk --- which in weakly magnetized disks is typically a cyclic, dynamo-like process \citep{brandenburg95,stone96,davis10,oneill11,simon12} --- can transport substantial amounts of energy vertically, and could heat low-density gas to create a corona. Turbulent vertical energy transport, moreover, has a stabilizing influence on thermal instability \citep{zhu14}. It is at best unclear, however, whether a disk model that sandwiches a gas or radiation-pressure dominated mid-plane with a magnetically dominated corona can undergo a state transition. Fully converting a thin, optically thick accretion disk to a tenous hot state requires such a dramatic decrease in density that it is hard to see how it could be accomplished without some catastrophic event to destroy the disk, and such events appear incompatible with observations that suggest rather seamless transitions among states \citep{dunn10, kalemci13}. 

In this paper, we suggest that the full catalog of XRB accretion states can be understood in terms of a competition between the dynamo generation of toroidal field and its buoyant rise, provided that the disk mid-plane is magnetically dominated. Magnetically dominated disk solutions can be constructed in a manner that parallels the standard Shakura-Sunyaev (1973) models \citep{pariev03}, though it is not obvious analytically what sets the level of magnetization. \cite{begelman07} presented models for maximally magnetized, magnetically supported disks, in the limit where radiation pressure dominates over gas pressure. The maximal level of magnetization was chosen to satisfy the marginal criterion for axisymmetric MRI to operate in a disk with a suprathermal toroidal magnetic field ($p_{\rm B} > p_{\rm gas}$) and a weak poloidal field, $v_{\rm A}^2 < c_{\rm g} v_{\rm K}$, where $v_{\rm A}$ is the Alfv\'en speed associated with the toroidal field, $c_{\rm g}$ is the sound speed associated with the gas pressure only, and $v_{\rm K}$ is the Keplerian speed \citep{pessah05}. Magnetically dominated disks are known to have various phenomenologically interesting properties. They are stable against thermal and viscous instability \citep{begelman07}, apparently persistent if they can be established in the first place \citep{johansen08}, and resistant to fragmentation \citep{begelman07,gaburov12}. 

Numerical simulations show that a magnetically dominated disk forms whenever the poloidal flux exceeds a threshold value \citep{bai13}, given by the requirement that the pressure in the poloidal field reaches about a percent of the sum of the gas and radiation pressures.  When this condition is met, the toroidal field becomes the dominant form of pressure throughout the disk, supporting the gas against the vertical component of gravity. Here we study the implications of the resultant vertical structure for XRB states. The key result is that magnetically dominated XRB disks may possess a single MRI-active layer, centered on the equator, or two active layers separated by a dead zone in which the \cite{pessah05} criterion for MRI activity is violated. When a dead zone exists, substantial amounts of magnetic energy can be injected into the gas at the very low densities where electrons and ions are thermally decoupled, producing a spectrally hard flux component that can co-exist with a cool equatorial region over a range of accretion rates. In its simplest form, the state of the disk is a local function of the accretion rate and $\alpha$, which is itself determined by the strength of the net poloidal flux. Following \cite{begelman14}, we appeal to global changes in the net flux to explain the hysteresis seen in transitions between the states.

In subsequent sections, we present a simple analytical model for the vertical structure of accretion disk+corona systems, and apply this stratified model to XRBs.  In Section 2 we show that the competition between toroidal field generation and escape is finely balanced, with small changes in dynamo and buoyancy parameters leading to large changes in the ratio of magnetic pressure to gas+radiation pressure near the equator.  We estimate the gradients of magnetic pressure and density, both when MRI is active and when it is suppressed.  In Section~3 we use this model to extract the basic phenomenology of magnetically supported disks, including the likely suppression of MRI in regions where the \cite{pessah05}  criterion is violated.  We apply these results to XRBs in Section~4, where we argue that the relative amounts of energy released in the lower and upper MRI zones determine whether the flow is in a soft, intermediate or hard state.  (We note that \cite{rozanska15} arrive at a similar vertical structure, that includes a dead zone, from spectral fitting.) By linking the vertical structure to the outcome of dynamo activity in magnetically supported disks, our model provides an explanation for the rich phenomenology of XRB state transitions.  We discuss the results and conclude in Section~5.

\section{Toroidal field generation and escape}

Consider a steady-state disk atmosphere at cylindrical radius $R$ in the gravitational field of a central mass $M$, supported vertically by some combination of thermal+radiation pressure, $p(z)$, and magnetic pressure due to a toroidal field, $p_{\rm B} (z) = B_\phi^2/8\pi$.  Denoting the vertical gravity by $g_z \approx GMz/R^3 = \Omega^2 z$, we have the equation of hydrostatic equilibrium,
\begin{equation}
\label{hydrostat1}
 {d\over dz} (p + p_{\rm B}) =  - \rho \Omega^2 z,
\end{equation} 
where $\rho$ is the density.  A second equation is provided by considering the vertical flux of magnetic energy (Poynting flux), which we treat phenomenologically according to a set of physically motivated assumptions about its creation and escape. Specifically, we suppose that MRI liberates accretion energy {\it locally} at a rate (per unit volume) $\alpha \Omega (p + p_{\rm B})$, of which a fraction $\alpha_{\rm B}/\alpha$ is channeled into the toroidal field.  This toroidal field escapes, via the Parker instability, with a mean vertical speed $v_{\rm B} (z)$, and thus carries a mean Poynting flux $F_{\rm B}(z) = 2 p_{\rm B} v_{\rm B}$.  Because the configuration is expected to be far from marginal stability, we expect the rise of field lines to reflect the near free-fall of gas sliding along buckled field lines, and thus take 
\begin{equation}
\label{vb}
v_{\rm B}(z) = \eta \Omega z , 
\end{equation}
where $\Omega z$ is the local free-fall speed at $z$ and $\eta < 1$ accounts for geometric effects, such as field-line tangling, finite time scale of reconnection, etc., that might impede the overall energy transport.  We assume (at least initially) that $\eta$ is independent of $z$ but it need not be a universal constant.

Rather than lifting the gas uniformly, the field repeatedly buckles (probably on a transverse scale larger than the local height: Tout \& Pringle 1992), allowing matter to slide downward as it rises.  Large-scale toroidal loops of field then reassemble themselves slightly higher in the disk atmosphere, through reconnection, and the process repeats.  This process both slows down the net rise of the field, compared to the instability time scale, and causes the magnetic field to lose energy to the gas as it rises, at a rate (per unit volume) $-v_{\rm B} dp_{\rm B}/dz$.  Some of this energy goes initially into potential energy of the gas, then kinetic energy as the gas drops back, and ultimately internal energy as the kinetic energy is dissipated.  Additionally, magnetic energy can be converted directed into kinetic energy and heat at the reconnection sites.  We gloss over these details and assume that all of the energy lost by the Poynting flux goes into heating the gas locally.  An analogous heating prescription is frequently used in treatments of cosmic-ray transport in which cosmic rays stream through the background plasma but are isotropized in a frame co-moving with an ensemble of Alfv\'en waves that scatter the cosmic rays \citep{skilling75}.  In this case, the cosmic-ray streaming speed (basically a weighted average wave speed) plays the role of $v_{\rm B}$.  An equivalent description was used to describe ``effervescent heating" of the intracluster medium of a cluster of galaxies by buoyant bubbles of hot plasma injected by a central radio galaxy \citep{begelman01,ruszkowski02}. 

If there are no additional energy losses, we may write down the following equation for the Poynting flux,
\begin{equation}
\label{Poynting1}
{d F_{\rm B}\over dz} = 2 v_{\rm B} {dp_{\rm B}\over dz} + 2 p_{\rm B} {dv_{\rm B}\over dz} =  v_{\rm B} {dp_{\rm B}\over dz} + \alpha_{\rm B} \Omega (p + p_{\rm B}),
\end{equation}
which simplifies to
\begin{equation}
\label{Poynting2}
{d \over dz}(p_{\rm B} v_{\rm B}^2) =  \alpha_{\rm B} \Omega v_{\rm B}(p + p_{\rm B}).
\end{equation}

Defining the central values of pressure and density, $p_0 = p(0)$, $p_{\rm B0} = p_{\rm B} (0) \equiv \beta_0^{-1} p_0 $, $\rho_0 = \rho (0)$, a gas pressure scale height $H \equiv \Omega^{-1}(p_0/ \rho_0)^{1/2}$, and a dimensionless vertical variable
\begin{equation}
\label{ydef}
y \equiv {z^2\over 2 H^2},
\end{equation}
and using eq.~(\ref{vb}) for $v_{\rm B}$, we obtain the dimensionless equations of hydrostatic equilibrium and magnetic energy,
\begin{equation}
\label{hydrostat2}
{d\over dy} (\tilde p_{\rm B} + \beta_0 \tilde p) = - \beta_0 \tilde\rho
\end{equation}
\begin{equation}
\label{Poynting3}
{d\over dy} (\tilde p_{\rm B} y) = {\alpha_{\rm B} \over 2 \eta} (\tilde p_{\rm B}+ \beta_0 \tilde p ),
\end{equation}
where a tilde denotes a quantity normalized to its equatorial value, so that $\tilde p_{\rm B}(0) = \tilde p (0) = \tilde\rho(0) = 1$.

Since equation (\ref{Poynting3}) is singular at $y = 0$, the parameters must satisfy a regularity condition to ensure that $\tilde p_{\rm B}' (0)$ is finite (where a prime indicates differentiation with respect to $y$),
\begin{equation}
\label{betacond}
\beta_0 = {2\eta\over \alpha_{\rm B}} - 1.
\end{equation}
We can therefore write eq.~(\ref{Poynting3}) as 
\begin{equation}
\label{Poynting4}
y \tilde p_{\rm B}' = {\beta_0 \over 1 + \beta_0} (\tilde p - \tilde p_{\rm B}).
\end{equation}
Equation (\ref{betacond}) reveals an important generic feature of the model: the ratio of gas+radiation pressure to magnetic pressure on the equator, $\beta_0$, depends solely on the ratio of the escape speed parameter $\eta$ to the flux production parameter $\alpha_B$.  When this ratio is $\gg 1$, flux can readily escape as it is produced and gas+radiation pressure dominates the disk.  When this ratio equals 1, the pressures are equal, but only a small decrease in the ratio, to 1/2, drives the ratio of gas+radiation pressure to magnetic pressure formally to zero.  Thus, an order unity change in a simple parameter ratio makes an enormous difference to the physical nature of the disk.  

To determine the pressures as a function of height requires a third equation involving the density, such as an equation of state or a radiative diffusion equation.  However, some generic features of the pressure profile are insensitive to the detailed behavior of the density and can be gleaned from the asymptotic behavior of solutions at small and large heights above the equator.  For example, by assuming that $\tilde\rho = 1 + O(y)$ for $y\ll 1$, we find that the normalized pressures close to the equator are given by
\begin{equation}
\label{pressure1}
\tilde p_{\rm B} \approx 1 - {\beta_0 \over 2 (1 + \beta_0)}y; \ \ \ 
\tilde p \approx 1 - {(1 + 2\beta_0) \over 2 (1 + \beta_0)}y.
\end{equation}
This implies that $\tilde p_{\rm B} - \tilde p \approx y/2$ increases with height, at least close to the equator.  In fact, we can prove that $\tilde p_{\rm B}/\tilde p > 1$ at all heights, by combining equations (\ref{hydrostat2}) and (\ref{Poynting4}) to yield
\begin{equation}
\label{dominance}
(1 + \beta_0)  {d \ln( \tilde p_{\rm B}/ \tilde p )\over d\ln y}  = {\tilde\rho \over \tilde p} + \beta_0 \left( {\tilde p\over \tilde p_{\rm B}} - 1 \right)  + 1 - {\tilde p_{\rm B} \over \tilde p}.
\end{equation}
Since the right-hand side of eq.~(\ref{dominance}) is positive definite for all $\tilde p_{\rm B} \leq \tilde p$, we see that the ratio always tends to values $> 1$.
\begin{figure*}
\centering
 \includegraphics[ width=3.2in,height=2.20in,keepaspectratio]{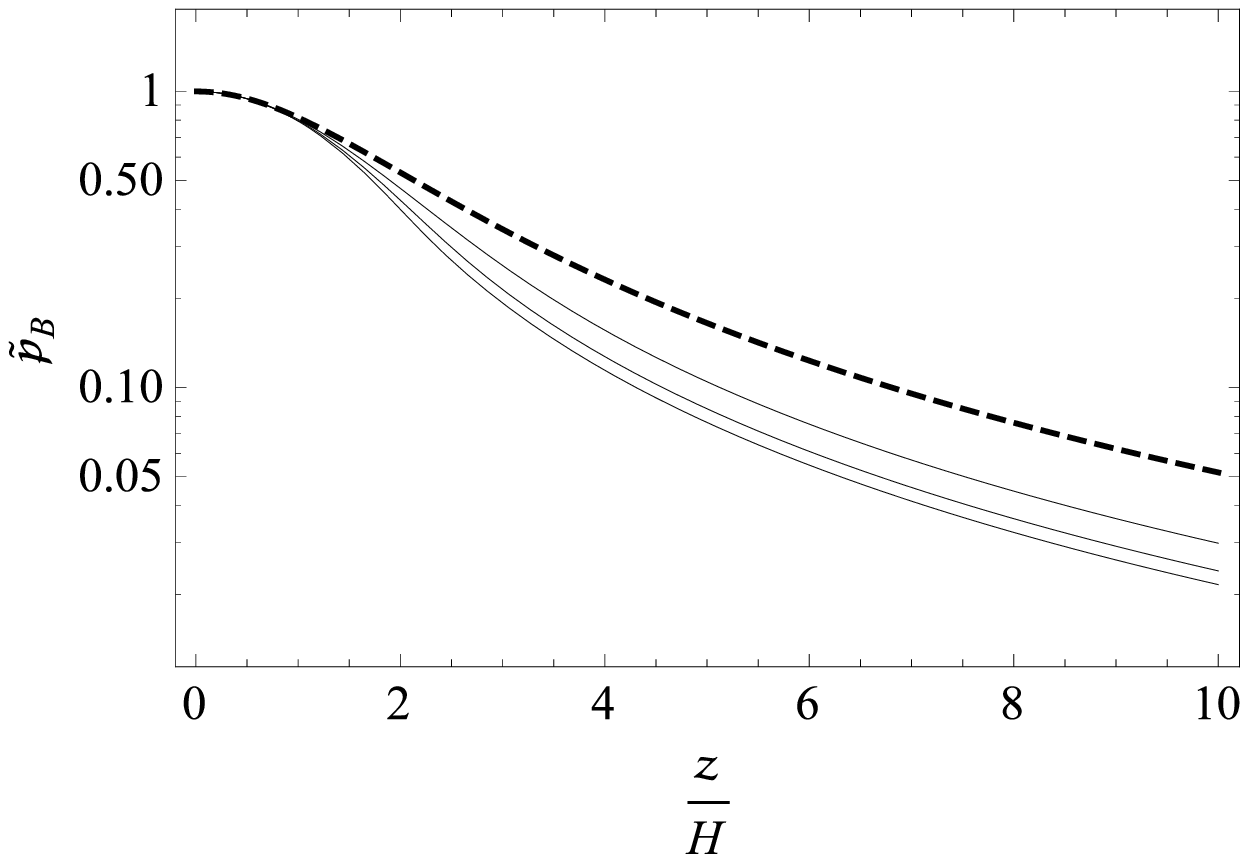}
 \includegraphics[ width=3.20in,height=2.20in,keepaspectratio]{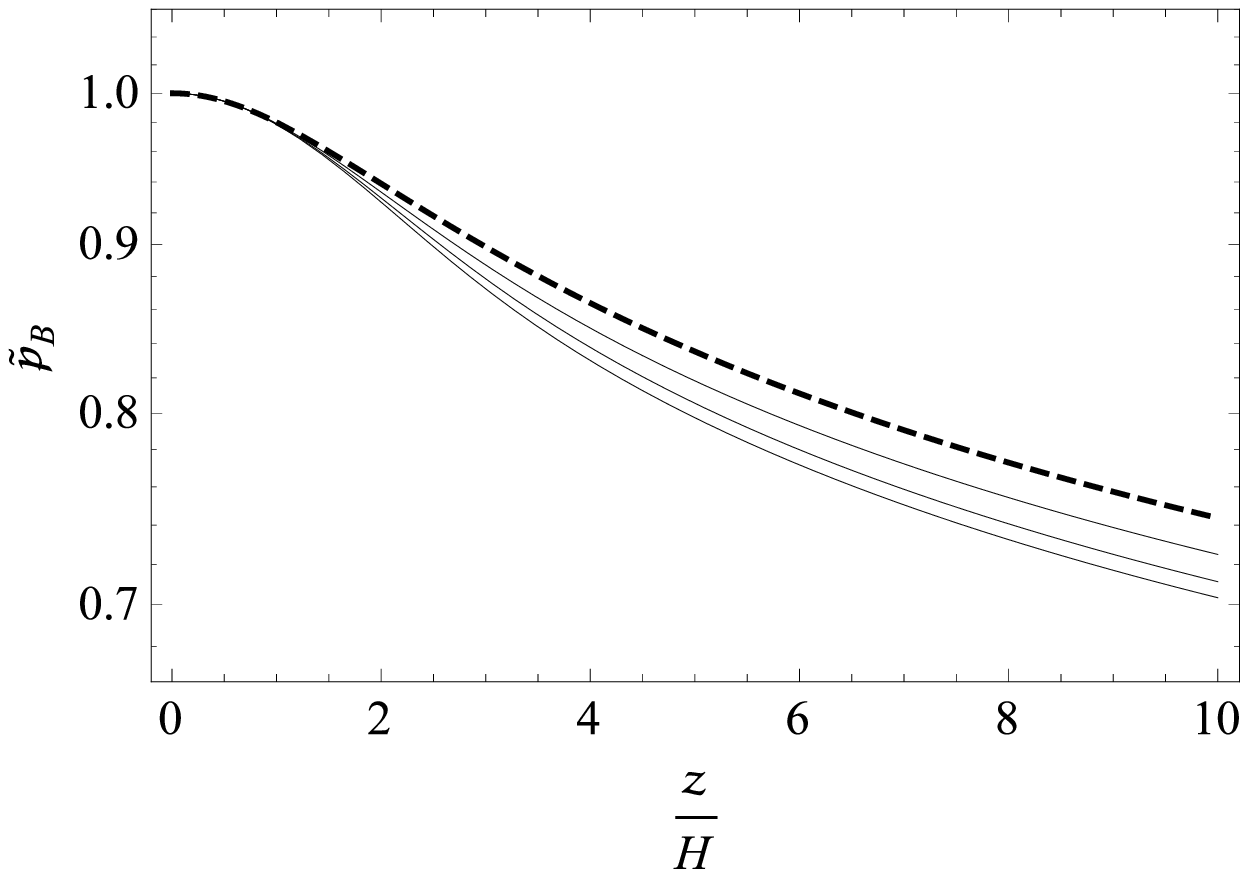}
  \caption{Normalized magnetic pressure $\tilde p_{\rm B}$ from solutions of equations (\ref{hydrostat2}) and (\ref{Poynting4}) for an equation of state $\tilde p = \tilde\rho^{\gamma}$, compared with the analytic approximation (\ref{bpinterp1}). Ordering of solid curves (top to bottom, along right-hand side) corresponds to $\gamma=1, 4/3, 5/3$. Dashed curve is analytic approximation. Left: Gas+radiation pressure-dominated case, $\beta_0=10$.  Right: Magnetic pressure-dominated case, $\beta_0=0.1$. }
  \label{fig:Fig1XRB}
\end{figure*}

If we assume that $\tilde p_{\rm B}\gg \tilde p$ at large $y$, we obtain the asymptotic scaling $\tilde p_{\rm B} \propto y^{-\beta_0/(1+\beta_0)}$, which we can combine with eq.~(\ref{pressure1}) to obtain a single expression that interpolates between the two limits,
\begin{equation}
\label{bpinterp1}
\tilde p_{\rm B} \approx \left( 1 + {y\over 2} \right)^{-\beta_0/(1+\beta_0)}.
\end{equation}
For a polytropic equation of state, $\tilde p = \tilde\rho^\gamma$, eq.~(\ref{bpinterp1}) (which does not depend on $\gamma$) gives an extremely good fit to the normalization as well as the slope of $\tilde p_{\rm B}$ at large $y$ (Fig.~1), implying that the magnetic pressure is insensitive to the gas equation of state.  Not surprisingly, the gas pressure, and thus the density, does depend on the equation of state (Fig.~2). 
\begin{figure*} 
\centering
 \includegraphics[width=3.20in,height=2.20in,keepaspectratio]{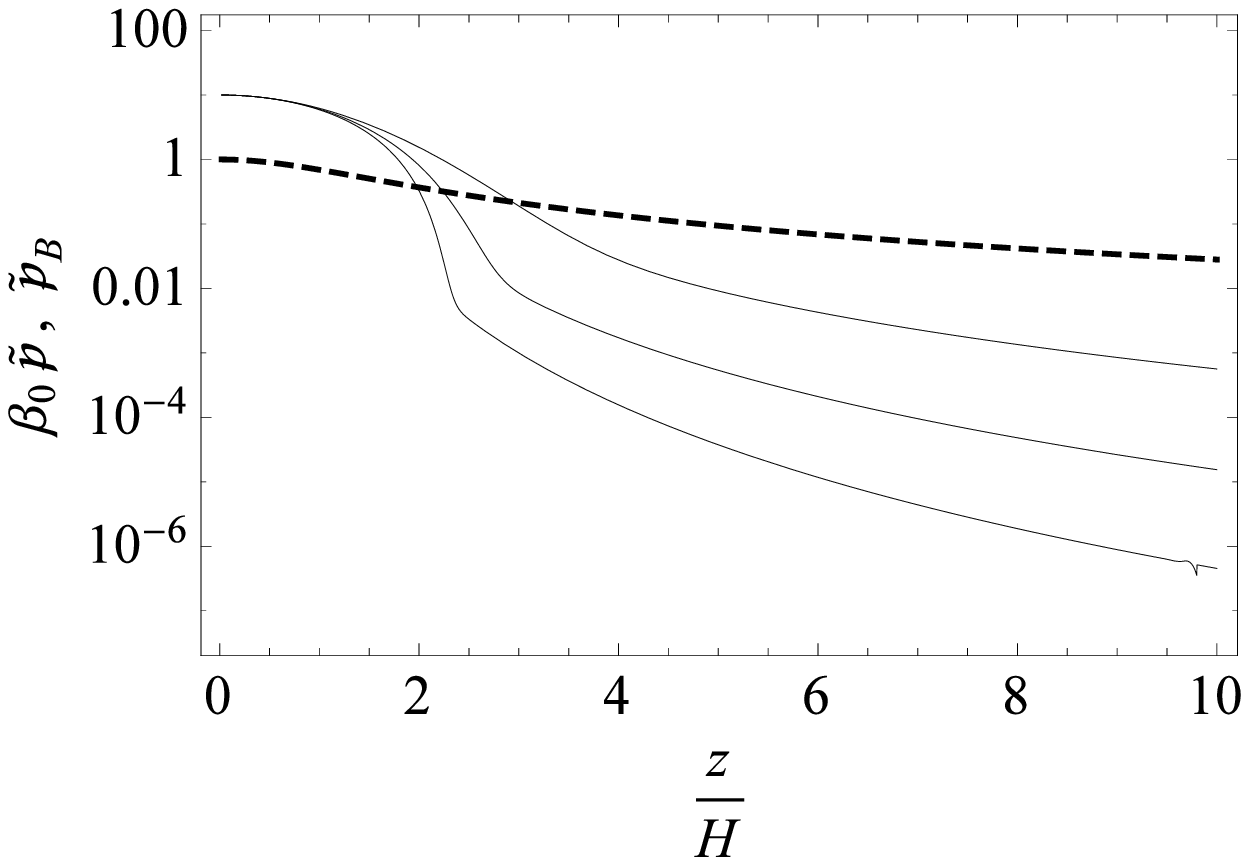}
 \includegraphics[width=3.30in,height=2.20in,keepaspectratio]{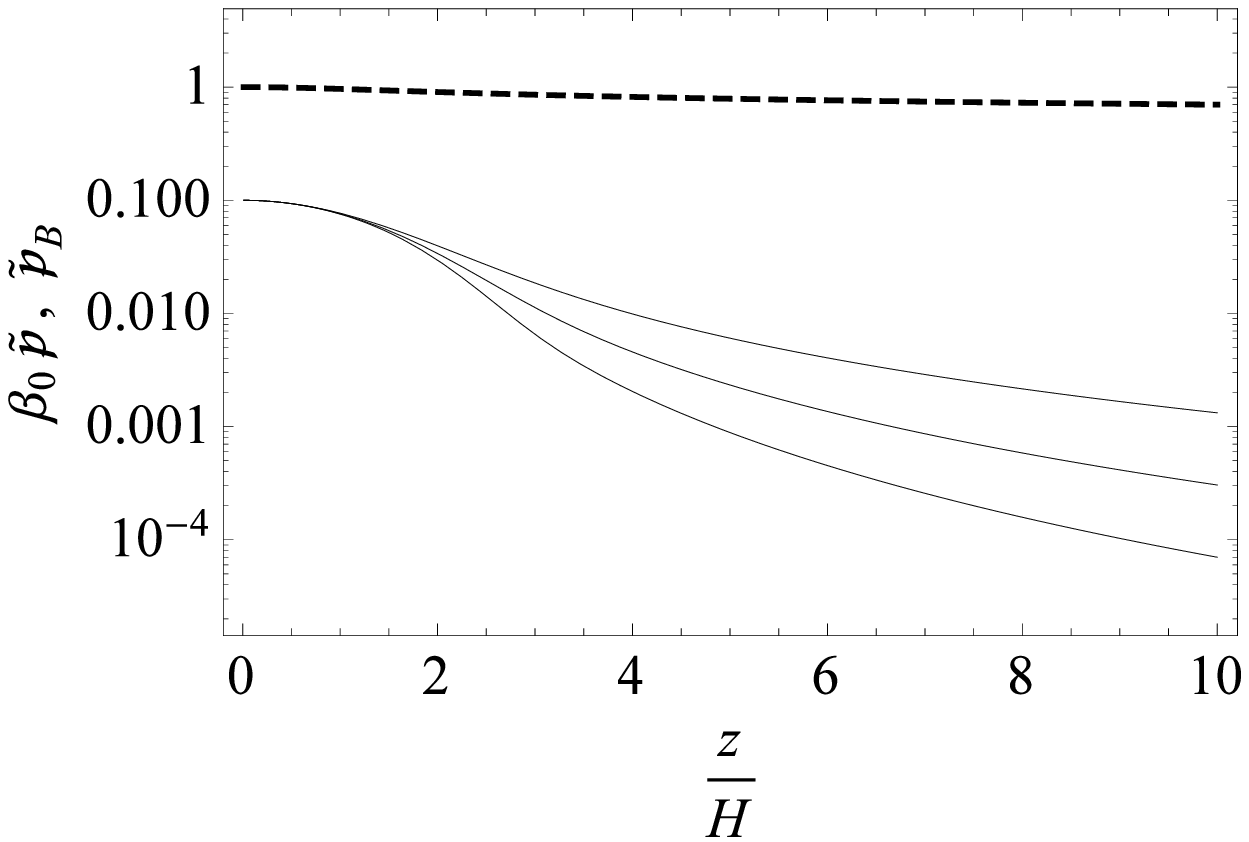}
  \caption{Gas+radiation pressure (solid lines) relative to normalized magnetic pressure (analytic approximation (\ref{bpinterp1}): dashed lines) as a function of height, for an equation of state $\tilde p = \tilde\rho^{\gamma}$. Left: Case where gas+radiation pressure dominates on the equator, $\beta_0=10$.  Right: Magnetic pressure-dominated case, $\beta_0=0.1$.  Ordering of solid curves (top to bottom) corresponds to $\gamma=1, 4/3, 5/3$.  Even when gas+radiation pressure dominates on the equator, magnetic pressure becomes dominant within a few scale heights.}
  \label{fig:Fig2XRB}
\end{figure*}
 
For polytropic equations of state, magnetic pressure dominates above a few scale heights (Fig.~2), even when gas+radiation pressure is dominant on the equator.  In fact, for these models the ``temperature" $p/\rho$ decreases with height, whereas the effective virial temperature increases $\propto z^2$.  Thus, these coronal models are ``cold" and fully supported by the gradient of magnetic pressure.  

A remarkable feature of these models is that relative importance of coronal vs.~disk heating depends entirely on the ratio of gas+radiation pressure to magnetic pressure on the equator, $\beta_0$.  The vertically integrated heating rate from the equator up to $z \gg H$ converges to a constant for a gas+radiation dominated disk, $\beta_0 > 1$, whereas it continues to increase $\propto \tilde p_{\rm B} z \propto z^{(1 - \beta_0)/(1 + \beta_0)}$ for $\beta_0 < 1$.  Thus, in this model the sensitive switch, eq.~(\ref{betacond}), that determines $\beta_0$, also determines whether most of the dynamo energy is dissipated below the disk photosphere, or travels far above the disk plane into a region where optical depths are low and cooling may be inefficient.    

Frequent reversals of the toroidal field, which are observed in accretion disk simulations with a net magnetic flux \citep{davis10,oneill11,simon12} can steepen the magnetic pressure gradient at large $z$, due to increased magnetic reconnection across the rising current sheets that separate the reversed regions.  In an Appendix, we show how to incorporate such reversals into our model.  However, the dynamos producing magnetically supported ($\beta_0 < 1$) disks appear to reverse infrequently, if at all, \citep[][Salvesen et al., in preparation]{bai13}, allowing us to apply models with a unidirectional field.  

\section{Phenomenology of magnetically supported disks}

In this section we extract what we believe are the salient lessons about the generic behavior of magnetically supported disks and coronae, which will form a basis for our applications to X-ray binary state transitions.  Since we are concerned here with magnetically supported disks that exhibit infrequent toroidal field reversals \citep{bai13}, we will ignore the effects of reconnection losses across large-scale current sheets.

\subsection{When are disks magnetically dominated?}

We have shown that the degree of magnetic domination can be described by the difference between two parameters, $2 \eta - \alpha_{\rm B}$, where $\alpha_{\rm B}$ describes the rate of field creation and $\eta$ describes its rate of loss through buoyancy.  Obviously these parameters cannot be physically independent, since if $\alpha_{\rm B}$ approached $2 \eta$ the magnetic pressure would become infinitely strong compared to other forms of pressure.  Before this limit could be reached, additional field would be expelled, or the dynamo quenched, leaving $2\eta$ at least slightly larger than $\alpha_{\rm B}$.  

Simulations suggest that the difference between these parameters is governed by the poloidal $\beta-$parameter $\beta_{\rm p}$, i.e., the ratio of the gas+radiation pressure to the pressure in the poloidal components of the magnetic field (Bai \& Stone 2013, Salvesen et al., in preparation).  This parameter is also thought to govern the effective value of the viscous $\alpha-$parameter, with $\alpha$ reaching values $\sim O(1)$ for $\beta_{\rm p} \ltorder 10^2$, while settling at much lower values $\sim 0.03$ for $\beta_{\rm p} \gtorder 10^4$.  Similarly, the toroidal magnetic pressure associated with the dynamo is very small for large values of $\beta_{\rm p}$, begins to approach the gas+radiation pressure near the equator for $\beta_{\rm p} \ltorder 10^3$, and becomes dominant throughout the disk (by about a factor $\sim 10$) for $\beta_{\rm p} \sim 10^2$. These poloidal field strengths fall squarely in the regime where local numerical simulations predict that magneto-centrifugal winds would be launched \citep{blandford82,suzuki09}, though the local approximation is ill-suited to determining the resultant mass and angular momentum loss rates \citep{fromang14}. Wind-driven flux transport  would assist in the maintenance of the relatively strong poloidal fields needed for our model, though the details of such transport would need a global treatment that we do not attempt here.

For the remainder of this paper we will assume that the strength of the dynamo, hence the dominance of the toroidal magnetic field, is set by some external parameter, presumably $\beta_{\rm p}$.  We will characterize the dominance of the toroidal pressure in terms of an equatorial value of the toroidal $\beta-$parameter, $\beta_0$. 

\subsection{Approximate vertical structure}

In regions where MRI is active and the disk is magnetically supported, we will assume that the magnetic pressure above $z \sim H$ varies as a weak power of $z$, $p_{\rm B} \propto z^{-2\beta_0}$, and therefore can be treated roughly as a constant for $\beta_0 \ll 1$.  Maintenance of approximate hydrostatic equilibrium then demands that $\rho \propto z^{-2}$.  

Where MRI is suppressed, the gradient of magnetic pressure is governed by losses in the vertical Poynting flux, according to eq.~(\ref{Poynting2}) with $\alpha_{\rm B} = 0$.  We then have $p_{\rm B} \propto z^{-2}$ and $\rho \propto z^{-4}$.  

In such ``dead zones," the Alfv\'en speed tracks the local free-fall speed, $v_{\rm A}^2 = 2 p_{\rm B}/\rho = \Omega^2 z^2$.  In MRI-active regions, however, the weak pressure gradient implies that the local Alfv\'en speed exceeds the free-fall speed by a factor $\sim \beta_0^{-1}$. Taken at face value, these scalings would imply a sharp density inversion at the transition from an MRI-active zone to a dead zone, but we expect such an inversion to be unstable, and to be smoothed, e.g., by a local increase in the buoyant rise speed or a gradual rather than sudden drop in $\alpha_{\rm B}$.  For purposes of outlining our state transition model, however, we will assume the simple piecewise scalings outlined above.      

\subsection{Suppression of MRI}

In order for MRI to operate in the presence of both poloidal and toroidal magnetic field, two conditions must be satisfied. First, the poloidal magnetic pressure must be smaller than the pressure responsible for setting the scale height, which in this case means that the poloidal field must be smaller than the toroidal field.  We assume that this is always satisfied near the disk midplane, since the presence of a moderate poloidal field appears to be capable of catalyzing the creation of a much stronger toroidal field through dynamo action.  At large heights, however, this condition might not be met if the toroidal magnetic pressure drops significantly across a dead zone, since poloidal field strength should be roughly independent of $z$.

The second condition provides an upper limit to the toroidal field.  Pessah \& Psaltis (2005) showed that MRI is suppressed, in the presence of a strong toroidal field and weak poloidal field, when 
\begin{equation}
\label{MRIsupp}
v_{\rm A}^2 > c_{\rm g} v_{\rm K}
\end{equation}
where $c_{\rm g}$ is the thermal (gas) sound speed and $v_{\rm K}$ is the Keplerian speed.  Begelman \& Pringle (2007) interpreted this criterion in terms of accretion disk structure, and argued that the sound speed should correspond to that of the gas only and should not include contributions from radiation pressure because of the highly diffusive nature of radiative transport across small scales. They also noted that this criterion is consistent with earlier results on non-axisymmetric MRI for a purely toroidal field \citep{terquem96}. We are not aware of any numerical simulations that have directly tested the Pessah \& Psaltis (2005) criterion, though the results of \cite{gaburov12} show that magnetically dominated disks are both relatively stable and can support MRI in the supra-thermal disk core. The \cite{gaburov12} simulations, however, did not include the extra physics (such as the possibility of separate electron and ion temperatures) that we argue is essential for a model of XRB state transitions.

\section{State transitions in X-ray binaries}

\begin{figure}
\vspace{-1.2truein}
\includegraphics[width=0.5\textwidth]{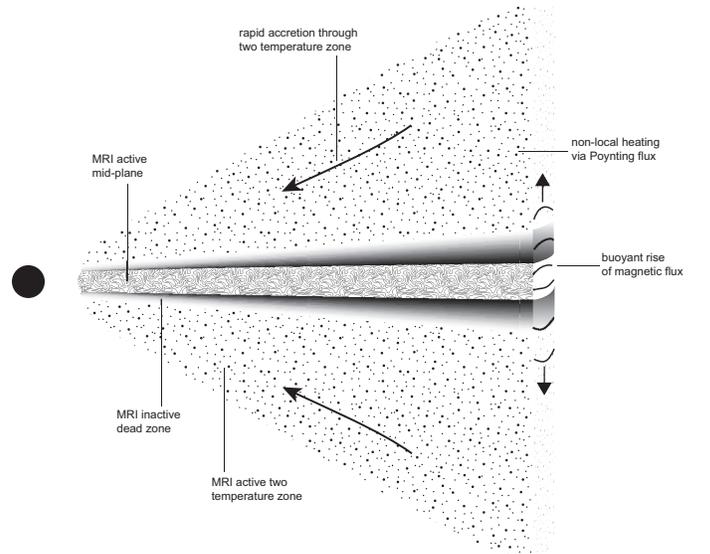}
\caption{Illustration of the vertical structure of a magnetically dominated disk in the regime which we identify with 
the intermediate state of XRBs. A strongly magnetized MRI-active mid-plane is separated from an MRI-active two temperature 
corona by a dead zone where the toroidal magnetic field is too strong to admit instability.}
\label{fig_two_temperature}
\end{figure}

\subsection{Schematic approach to state transitions}

We consider a magnetically supported disk-corona system that is MRI-active close to the equator, and follow its behavior with increasing height.  Since the $v_{\rm A}^2 \sim z^2$, eq.~(\ref{MRIsupp}) will be satisfied above some height $z_1$, shutting off MRI, unless $c_{\rm g} \propto T^{1/2}$ increases very rapidly with height, where $T$ is the temperature.  In the optically thick, thermalized equatorial region of an accretion disk, $T$ is a relatively weak function of height, and therefore the system enters a dead (MRI-inactive) zone at $z > z_1$.  

In the dead zone, gas continues to be heated by the rising toroidal flux from below, but at a rate that decreases with height $\sim p_{\rm B}\propto z^{-2}$ .  $v_{\rm A}^2$ continues to increase $\propto z^2$ (albeit with a slightly different normalization), but the most dramatic change is the steep decline in density, $\rho \propto z^{-4}$.  For the case of X-ray binaries (XRBs), we will argue that the decline in the cooling rate, due to the decrease in density, far outweighs the weakening heating rate and leads to the thermal decoupling of electrons and ions and a runaway increase of ion temperature, $T_{\rm i} \gg T_{\rm e}$.  Remarkably, we find that $c_{\rm g} \propto T_{\rm i}^{1/2}$ can increase faster than $z^2$, leading to the re-establishment of MRI above a height $z_2$, provided that the poloidal magnetic field is not too strong.  We note that the energy source for this second MRI zone is not the Poynting flux generated near the equator, but rather the local differential rotation of the flow within the zone, and the gravitational binding energy liberated by accretion of this gas.  However, the Poynting flux from below is crucial for catalyzing the re-establishment of MRI by heating the gas sufficiently to drive it back into the MRI-unstable regime.  Once established, this upper MRI zone extends all the way to the top of the corona at $z \sim R$.  

Because $p_{\rm B}$ is roughly constant with height within each MRI-active zone, the total dissipation within that zone is $\sim \alpha \Omega p_{\rm B} (z_{\rm max} - z_{\rm min})$, where $z_{\rm min}$ and $z_{\rm max}$ are the lower and upper boundaries.  For simplicity in the approximate expressions derived below, we will assume $z_{\rm max}\gg z_{\rm min}$. Then, given that an upper MRI zone exists, $z_2 < R$, we can estimate the ratios of accretion rates and powers associated with the two MRI-active zones:  
\begin{equation}
\label{MRIratio}
{\dot M_2 \over \dot M_1} = {L_2\over L_1} \approx {p_{\rm B2} R \over p_{\rm B1} z_1} \sim {z_1 R \over z_2^2},
\end{equation}
where the final relation reflects our extremely crude approximation that $p_{\rm B2}/ p_{\rm B1} \sim (z_1 / z_2)^2$. In this 
picture the two MRI-active zones share the same poloidal flux, but are otherwise independent. While we remain in a regime 
where the dominant toroidal field controls MRI-activity, this is locally a reasonable assumption. Globally, however, neighboring 
regions of the disk will be coupled by the evolution of the poloidal field, which will be qualitatively different depending on 
whether the vertically extended upper MRI-active layer is present or not \citep{lovelace09}. We expect inward transport of the 
net poloidal flux to be greatly facilitated when there is a fast-accreting upper layer, though the details will depend substantially 
on the strength and nature of MHD outflows which this geometry also favors.

Schematically, we associate the low/hard, intermediate/hard, intermediate/soft and high/soft accretion states with the cases $\dot M_2 / \dot M_1 \gg 1$, $\gtorder 1$, $\ltorder 1$, and 0, respectively.  In the following section we estimate these criteria quantitatively for conditions appropriate to X-ray binaries, and flesh out some of the associated observational characteristics. 

\subsection{Soft state}

The structure of thin, magnetically supported disks with radiation pressure exceeding gas pressure is discussed at length in \cite{begelman07}, under the assumption that the disk marginally satisfies the instability criterion of \cite{pessah05}.  Readers are referred to that paper for insight into results presented in this section.  As we saw earlier, the vertical structure of a magnetically dominated disk guarantees that most of the accretion will occur close to marginal (maximally magnetized) conditions for MRI, although the optically thick disk interior may hide a large column density of slowly accreting matter. We denote the dimensionless accretion rate through the equatorial layer as $\dot m_1 = \dot M_1/\dot M_{\rm Edd}$, where $\dot M_{\rm Edd} = L_{\rm Edd}/c^2 = 4\pi GM/ \kappa c$ is the Eddington accretion rate, without the often-used efficiency factor.  Here $M = 10 m_1 M_\odot$ is the mass of the central object and $\kappa$ is the opacity.  We scale the radius to $R_{\rm g} = GM/c^2$, $x = R/R_{\rm g}$. The upper boundary of this accretion layer, where MRI shuts off according to the \cite{pessah05} criterion, is then given by $z_1$ where
\begin{equation}
\label{z1}
{z_1 \over R} \sim 0.07  \alpha^{-1/18} \dot m_1^{1/9} m_1^{-1/18}  x^{1/36} 
\end{equation}
\citep{begelman07}.  

This result is based on the assumption of LTE (the validity of which is discussed in \cite{begelman07}), including thermal equilibrium between electrons and ions, $T_{\rm i} \approx T_{\rm e}$.  We can check the latter assumption by using a simple approximate expression for the volume energy transfer rate from ions to electrons, minimally given by Coulomb collisions:
\begin{equation}
\label{Guilbert}
{\cal H}_{\rm ie} \sim 36 \left( {m_{\rm e} \over m_{\rm p}} \right)^{5/2} \rho^2 c^3 \kappa \left( {c \over c_{\rm g}} \right) \left( {T_{\rm i} \over T_{\rm e}} \right)^{1/2} {T_{\rm i} - T_{\rm e}\over T_{\rm e}}  
\end{equation}
\citep{guilbert85}, where we have taken the Gaunt factor $\ln \Lambda$ in their original expression to be $\approx 25$.
     
Assuming that much of the direct MRI heating goes into ions and equating $\alpha \Omega p_{\rm B}$ to ${\cal H}_{\rm ie}$, we obtain
\begin{equation}
\label{Guilbert2}
\Delta \equiv \left( {T_{\rm i} \over T_{\rm e}} \right)^{1/2} {T_{\rm i} - T_{\rm e}\over T_{\rm e}} \sim 0.03  \alpha^{29/18} \dot m_1^{-2/9} m_1^{-7/18}  x^{-47/36},  
\end{equation}
implying that the species are well-coupled at $z \ltorder z_1$.

At $z \gtorder z_1$, the magnetic pressure goes from being roughly constant to declining $\propto z^{-2}$, while the density tends toward a steep decline $\propto z^{-4}$.  Heating by MRI is gone, but the dissipation of Poynting flux heats the gas at a rate $- v_{\rm B} (dp_{\rm B}/dz) \propto p_{\rm B}$.  Denoting conditions at $z_1$ by a subscript 1, assuming that the changes in density and pressure slopes are instantaneous, and ignoring changes in normalization of the pressure and density power-laws, so that $p_{\rm B} \sim p_{\rm B1} (z/ z_1)^{-2}$, etc., we obtain $\Delta (z) \sim  (c_{\rm g}/ c_{\rm g1})(z/ z_1)^6 \Delta_1$, implying that thermal coupling breaks down very quickly, i.e., at $z \gtorder 2 z_1$ at this crude level of approximation.  

Above the decoupling point but still in the dead zone, we can assume $T_{\rm i} \gg T_{\rm e}$ and write $c_{\rm g} / c \approx (kT_{\rm i}/ \mu c^2)^{1/2}$, where $\mu \approx 0.6 m_{\rm p}$ is the mass per particle.  Writing $\theta_{\rm e} = kT_{\rm e} / m_{\rm e} c^2$, we can express the thermal balance for ions in the form
\begin{equation}
\label{Guilbert3}
{c_{\rm g} \over c} \sim 0.01 \theta_{\rm e}^{3/4} \left( { z \over z_1} \right)^3  \alpha^{31/36} \dot m_1^{-2/9} m_1^{-5/36} x^{-31/72}.  
\end{equation}
Now consider the criterion for the onset of MRI, $v_{\rm A}^2 < c_{\rm g} v_{\rm K}$.  If we make the reasonable assumption that $T_{\rm e}$ is increasing with $z$, then the right-hand side of this relation increases at least as rapidly as $z^3$.  But the quantity on the left-hand side is constrained, by the equation of hydrostatic equilibrium, to rise $\propto z^2$. Therefore, there is a chance that MRI will be re-established, above the dead zone, under conditions very much hotter and more tenuous than in the equatorial accretion zone.  
  
Denoting $z_2$ as the height where MRI is reestablished, we obtain
\begin{equation}
\label{z2}
{z_2 \over R} \sim 0.4  \left( { \theta_{\rm e} \over  \theta_{\rm e1}} \right)^{-3/4} \alpha^{-31/36} \dot m_1^{2/9} m_1^{5/36} x^{85/72}.  
\end{equation}
The quantity $ \theta_{\rm e1}$ depends on the temperature at $z_1$, where electrons and ions are still coupled.  A number of factors may affect the value of $\theta_{\rm e}$ at $z_2$, but to obtain a very rough estimate, we assume that the electrons close to $z_2$ are cooling by unsaturated Comptonization so that the Compton $y-$parameter $\ltorder 1$, depending on the amount of cooling required.  If we further assume that the electrons are subrelativistic and the scattering optical depth $\tau > 1$, then       
\begin{equation}
\label{ydef}
y \sim 4 \tau^2 \theta_{\rm e} \sim 4  \tau_1^2 \theta_{\rm e}\left( { z \over z_1} \right)^{-6} ,  
\end{equation}
where
\begin{equation}
\label{tau1}
\tau_1 \sim 300 \alpha^{-8/9} \dot m_1^{7/9} m_1^{1/9} x^{-5/9}  
\end{equation}
\citep{begelman07}.  We obtain
\begin{equation}
\label{thetae2}
\theta_{e2}\sim 0.03 y^{2/11} \alpha^{-73/99} \dot m_1^{20/99} m_1^{-1/99} x^{145/198}   
\end{equation} 
and
\begin{equation}
\label{z22}
{z_2 \over z_1} \sim 5 y^{-3/22} \alpha^{-83/198} \dot m_1^{29/99} m_1^{7/198} x^{-25/396}.  
\end{equation} 
Note the extreme insensitivity of this expression to all parameters except $\alpha$ and $\dot m$ --- in particular, this result hardly depends on the value of $y$, which is uncertain.  We also have
\begin{equation}
\label{z23}
{z_2 \over R} \sim 0.3 \alpha^{-0.5} \dot m_1^{0.4} ,  
\end{equation} 
where we have suppressed the very weak dependence on $y$, $m$, and $x$.  An upper, MRI-active layer can exist only if $z_2 / R < 1$.  If this layer does not exist, i.e., if  $\dot m > 20 \alpha^{1.2}$ at our extremely simplified level of approximation, then the flow is in the soft state.

\subsection{Intermediate states}

If $z_2 < R$ and both lower and upper MRI-active zones coexist, then the flow is in an intermediate state, i.e., the spectrum has both soft and hard components (Fig.~3).  Using eq.~(\ref{MRIratio}) to estimate the ratio of accretion rates, we have  
\begin{equation}
\label{MRIratio2}
{\dot m_2 \over \dot m_1} \sim 0.7 y^{0.3} \alpha^{0.9} \dot m_1^{-0.7} ,  
\end{equation} 
where we have suppressed the very weak dependence on $m$ and $x$, but have retained the dependence on $y$ for reasons that we will now explain.  If $\dot m_2 \ll \dot m_1$, i.e., most power emerging from the cool equatorial zone, then having $y \sim 1$ in the upper layer would overcool it.  Rather, we would expect $y \sim \dot m_2 / \dot m_1$, implying   
\begin{equation}
\label{softint}
\dot m_2 \sim 0.6 \alpha^{1.3}  
\end{equation} 
is roughly independent of $\dot m_1$ in this soft-intermediate state.  According to our simple scalings, the soft-intermediate state exists for $0.6 \alpha^{1.3} < \dot m < 20 \alpha^{1.2}$, i.e., potentially over a large range of luminosities.

When $\dot m \ltorder 0.6 \alpha^{1.3}$, $\dot m_2$ is comparable to or larger than $\dot m_1$, $y \sim O(1)$ is required in order to cool the upper layer, and we have 
\begin{equation}
\label{hardint}
\dot m_2 \sim 0.7 \alpha^{0.9} \dot m_1^{0.3} .   
\end{equation}    
Since the hard flux dominates over the soft flux, we identify this state as the hard-intermediate state.  

We add the caveat that we have assumed $\tau > 1$ and subrelativistic electrons at $z \sim R$ to obtain these expressions.  These assumptions may be violated, e.g., the optical depth in the upper radiating region may be $\ltorder 1$, implying that $y \propto \tau$ rather than $\tau^2$.  Whether the electrons become mildly relativistic in this limit depends on whether $y \ll 1$ (as in the soft-intermediate state) or $\sim 1$.  We note, however, that the hard spectral components in some XRBs in an intermediate state  show much higher high-energy cutoffs than ``pure" hard-state spectra \citep{motta09}. 

\subsection{Hard state}
 
When $\dot m \ll 0.6 \alpha^{1.3}$, $\dot m_2 \gg \dot m_1$ and there is plenty of energy available to evaporate the cool disk and incorporate its mass into the hot accretion flow.  We therefore suppose that the cool disk vanishes, and seek a solution for a single-zone, magnetically supported flow in which radiation pressure is negligible compared to gas pressure.  Equations (10)--(13) and (16) of \cite{begelman07} remain valid, while $p_{\rm g}$ is given by $\rho c_{\rm g}^2$.  Assuming Coulomb coupling between electrons and ions as before, and cooling of electrons by unsaturated Comptonization with $\tau > 1$ and $y \sim O(1)$, we find that $T_{\rm i} \gg T_{\rm e}$ for $\dot m > 10^{-2} y^{1/2} \alpha x$, with 
\begin{equation}
\label{hard1}
{c_{\rm g} \over c} \sim 0.5 y^{-3/7} \alpha^{-10/7} \dot m^{8/7} x^{-13/14},  
\end{equation}  
\begin{equation}
\label{hard2}
\theta_{\rm e}\sim 0.04 y^{1/7} \alpha^{-6/7} \dot m^{2/7} x^{1/7}.  
\end{equation} 
One can verify a posteriori that $\tau > 1$ for typical parameters, justifying use of the subrelativistic version of the $y-$parameter.  At lower values of $\dot m$ the electrons and ions would become thermally coupled, and thermalization of the radiation might occur.

Once the flow is in the hard state, it need not return to an intermediate state when $\dot m > 0.6 \alpha ^{1.6}$.  As $\dot m$ increases, both $c_{\rm g}/c$ and $\theta_{\rm e}$ are expected to increase.  Dynamical constraints make it likely that $c_{\rm g}/c$ will saturate at some value $\sim O(1)$, in which case the flow is essentially indistinguishable from a gas pressure supported flow.  If $\theta_{\rm e}$ also saturates at some value $\sim O(1)$, e.g., due to a steep increase in the cooling rate as the electrons become relativistic, then we recover the usual criterion for maintaining a hot (two-temperature) accretion flow, $\dot m /\alpha^2 <$ const. \citep{rees82}.  

\section{Discussion and Conclusions}

We have proposed that the spectral states in X-ray binaries arise as a consequence of the vertical structure of accretion disks whose dominant source of pressure support is a toroidal magnetic field.  As the field rises buoyantly, its tension reaches a level compared to local gas pressure at which MRI is quenched \citep{pessah05}.  However, the Poynting flux from the equatorial dynamo continues to heat the overlying gas, the density of which declines so steeply that it undergoes a steep temperature inversion.  This increase in gas sound speed can reignite MRI above the dead zone.  The existence of one or two MRI-active layers, carrying independent mass fluxes and producing various ratios of soft and hard radiation, accounts naturally for the rich phenomenology of soft, hard and intermediate states. The strong toroidal magnetic field protects both the soft and intermediate states from either thermal or viscous instability \citep{begelman07}.

Our model readily explains several features of XRB phenomenology that have proven hard to understand.  First, it explains the inevitability of soft-to-hard state transitions as the accretion rate declines.  In models where both cold and hot accretion states can exist at low accretion rates \citep{chen95,esin97}, the conversion of the former into the latter virtually requires explosive heating to drive the density down by several orders of magnitude, so that electrons and ions can decouple thermally.  This transition is usually imposed by fiat.  However, in our model the vertical stratification leads to a gradual transition, through soft- and hard-intermediate states, as the bulk of the energy generation passes smoothly from the cool layer to the hot layer with decreasing $\dot m$.  

Once the cool zone disappears, the hard state consists of a single MRI-active layer, dominated by gas pressure and with thermal properties rather different than those of the hard-intermediate state.  Because of the weaker density stratification when the dead zone is absent, both the ion and electron temperatures are somewhat lower than in the intermediate states (although $T_{\rm i}$ can still greatly exceed $T_{\rm e}$).  Thus we can explain the curious observational fact that the ``pure" hard state, where no thermal component is apparent, often has a lower high-energy cutoff than either of the intermediate states \citep{motta09}. Furthermore, because the accretion flow in the hard state is fundamentally different from that in the intermediate state, its transition back to the intermediate state during the rising phase of an outburst occurs at a different threshold.  This can help to explain the hysteretic cycle observed in state transitions.

The model discussed here predicts a local relationship between the mass accretion rate, $\alpha$, and the state of the disk (specifically, how much energy is dissipated near the equator versus low density regions higher up). We require that the disk pressure be dominated by toroidal magnetic field pressure at all heights, which in turn requires the presence of a poloidal field whose pressure exceeds about one percent of the gas plus radiation pressure. Observationally, the presence of jets during some phases of X-ray binary outbursts strongly suggests that poloidal field is present \citep[e.g.][]{blandford77,mckinney12}, but does not necessarily imply that it be of large scale \citep{parfrey15}. Disk winds, moreover, are inferred to be present in the soft and intermediate states of XRBs, consistent with a radially extended region that contains significant poloidal flux. 
Theoretically, maintaining even fairly weak poloidal fields in the inner disk is not necessarily straightforward; radial dragging of poloidal flux is inefficient in geometrically thin disks \citep{lubow94,guilet12}, and the extent to which an inverse cascade can generate locally strong fields from zero net flux initial conditions is unknown \citep{sorathia10,beckwith11}. We have previously suggested that the transition zone between geometrically thin and thick flows is a favorable site for stochastic flux accumulation \citep{begelman14}, and this mechanism may play a role in sustaining poloidal flux near the black hole. We note, in particular, that if the poloidal flux ever dropped below the level needed to sustain magnetic pressure dominance, the onset of thermal instability in radiation pressure dominated regions would create the geometry that is favorable for flux accumulation. More generally, changes in $\alpha$ driven by global changes in the poloidal flux appear necessary to explain why the intermediate state can exist at different levels of luminosity.

A complete model for state transitions must also account for the temporal properties of the states and their relationship to jet outflows.  Strong rms fluctuations on short timescales, and lower frequency QPOs, are key features of the hard and intermediate states.  In particular, the hard and hard-intermediate states exhibit Type C QPOs, which are strong and relatively coherent (with $Q-$values as large as $\sim 10$), with a range of frequencies ($\sim 0.1 - 10$ Hz) correlated with spectral hardness \citep{vignarca03,stiele13} and energy cutoff (E.~Kalemci, private communication); while the soft-intermediate states sometimes exhibit less coherent QPOs with roughly fixed frequencies ($\sim 6-8$ Hz) \citep{casella05, belloni11}.  The frequencies of these QPOs are all much lower than any dynamical or likely precession period in the region close to the black hole where most energy is generated, and we suggest that they reflect the quasi-periodic magnetic field reversals in the dynamo.  Simulations indicate that the intervals between flux reversals become quite long as the strength of the toroidal field increases, reaching tens of orbital periods, if not longer, for magnetically dominated flows (Bai \& Stone 2013; Salvesen et al., in preparation).  Such time scales are compatible with the observed QPOs, if they are produced at radii of $\sim 10-100 R_{\rm g}$.

The high coherence of Type C QPOs suggests that they are either global modes, or are produced at a specific point in the flow.  We note, however, that for the geometrically thick layers that likely modulate the QPOs, the inflow time scale is probably shorter than the field reversal time, implying that the polarity of the field is not produced locally, but rather propagates inward from some radius where the inflow time and field reversal time are comparable \citep{oneill11}.  For example, this could be the radius where the flow switches from a thin disk to a thick, magnetically supported state.     

The model presented here is highly schematic. Our crude approximations include a very simple treatment of the vertical structure (taking a uniform magnetic pressure in each MRI-active zone, using piecewise power-law approximations for the pressures and densities, parameterizing the critical question of how rapidly magnetic flux rises thorough the disk), and a minimal model for the electron-ion thermal coupling, which could differ from the Coulomb rate. The numerical values we have derived for the various thresholds are thus not expected to be accurate. We are encouraged, however, by the fact that given the reasonable assumption of a poloidal flux and resulting magnetic domination of the disk, the model shows promise for addressing a  wide range of XRB phenomena. Moreover, the most uncertain aspects --- the detailed vertical structure of strongly magnetized disks that include low density two-temperature regions, the accumulation and loss of poloidal flux in XRB disk geometries, and the variability of emission in the hard state --- appear amenable to study using local and global simulations that are feasible in the near-term.

\acknowledgements
We acknowledge the hospitality of the Department of Physics at the University of Oxford, the Institute of Astronomy at the University of Cambridge, the Institut d'Astrophysique de Paris, and the N.~Copernicus Astronomical Center in Warsaw, and thank Emrah Kalemci for extensive discussions on XRB observations.  Our work has been supported by NASA under Astrophysics Theory Program awards NNX11AE12G and NNX14AB42G. CSR additionally thanks the Simons Foundation and a Sackler Fellowship (Cambridge) for support.

\appendix\section{Effect of field reversals}

In many shearing-box simulations that show dynamo action, the direction of the toroidal field reverses quasi-periodically, on a time scale which can be shorter than the rise time of the field \citep{brandenburg95,stone96,davis10,oneill11,simon12,bai13,jiang14a}. Regions of oppositely directed toroidal field are separated by current sheets across which reconnection can take place, dissipating the magnetic energy more quickly than we have assumed in our analysis.  This extra dissipation could lead to a steeper magnetic pressure gradient and heating concentrated more equatorially than in the unidirectional model. Here we present an extremely simple model that attempts to capture the magnitude and vertical distribution of this dissipation in a time-averaged sense.  

Suppose that the toroidal flux undergoes periodic reversals, with a duty cycle $t_{\rm d} \sim (\xi \Omega)^{-1}$ and $\xi \ltorder 1$.  These reversals, which originate near the equator, launch rising current sheets across which reconnection takes place with a speed
\begin{equation}
\label{recon1}
v_{\rm rec} \sim \zeta v_{\rm A}, 
\end{equation}
where $v_{\rm A}$ is the Alfv\'en speed on either side of the current sheet.  The reconnection speed is defined as the speed with which the ambient magnetic field lines approach the current sheet from either side, and recent studies of collisionless magnetic reconnection suggest that $\zeta \sim 0.1-0.3$ \citep{zweibel09,uzdensky10}.  Thus, reconnection could occur rapidly compared to the rise time.  As one current sheet sweeps by, the heat deposited per unit volume (and the magnetic energy density correspondingly lost) is 
\begin{equation}
\label{recon2}
\Delta {\cal H} \sim  {2\zeta v_{\rm A} p_{\rm B} \over v_{\rm rise} }
\end{equation}
where $v_{\rm rise}$ is the rise speed of the current sheet, which need not be the same as the rise speed $v_{\rm B} \sim \eta \Omega z$ of the magnetic flux. Indeed, if the reconnection speed is required to be the same on both sides of the current sheet, then $v_{\rm rise}$ cannot equal $v_{\rm B}$; moreover, there may be additional bulk flows of matter, both up and down (since the problem is now time-dependent) that should be included as the passage of the current sheet deposits momentum into the flow. \citep[We suggest that these cyclic flows may be present in the density plot of Fig.~1 in][]{jiang14a}. These complications are well beyond the simple level of our model, and to avoid the proliferation of undetermined parameters we will simply assume $v_{\rm rise} \sim v_{\rm rec}$, which then allows $\zeta$ to cancel out of the problem. (We can subsume the uncertain validity of this assumption into the value of $\xi$.)  

The mean reconnection heating {\it rate} per unit volume, averaged over multiple cycles, is then   
\begin{equation}
\label{recon3}
{\cal H}_{\rm rec} \sim  2\xi \Omega p_{\rm B} . 
\end{equation}
Assuming that this energy comes from the Poynting flux generated by the MRI dynamo, we can replace eq.~(\ref {Poynting2}) by
\begin{equation}
\label{PoyntingRec1}
{d \over dz}(p_{\rm B} v_{\rm B}^2) =  \Omega v_{\rm B}[\alpha_{\rm B} p + (\alpha_{\rm B} - 2 \xi) p_{\rm B}].
\end{equation}
Using the same definitions for $y$ and normalized quantities as before, we find that eq.~(\ref{hydrostat2}) is unchanged, while the normalized magnetic energy flux equation (\ref{Poynting3}) is replaced by
\begin{equation}
\label{PoyntingRec2}
2 \eta {d\over dy} (\tilde p_{\rm B} y) = (\alpha_{\rm B} - 2\xi) \tilde p_{\rm B}+ \alpha_{\rm B}\beta_0 \tilde p 
\end{equation} 
and the regularity condition becomes
\begin{equation}
\label{betacondRec}
\beta_0 = {2(\eta + \xi)\over \alpha_{\rm B}} - 1.
\end{equation}
Defining a reconnection efficiency parameter, 
\begin{equation}
\label{nudef}
\nu \equiv {2 \xi \over \alpha_{\rm B}},
\end{equation}
we find that eq.~(\ref{Poynting4}) becomes 
\begin{equation}
\label{Poynting5}
y \tilde p_{\rm B}' = {\beta_0 \over 1 + \beta_0 - \nu} (\tilde p - \tilde p_{\rm B}).
\end{equation}
The analytic approximation for $\tilde p_{\rm B}$, eq.~(\ref{bpinterp1}), is then replaced by
\begin{equation}
\label{bpinterp1Rec}
\tilde p_{\rm B} \approx \left( 1 + {1 + \beta_0 - \nu \over 2(1+ \beta_0) -\nu } y\right)^{-\beta_0/(1+\beta_0 - \nu)}.
\end{equation}
Because the Poynting flux now has an additional dissipative channel, the magnetic pressure decreases more steeply with height.  The condition for magnetic pressure to dominate in the equatorial region ($\beta_0 <1$) is now $\eta + \xi < \alpha_{\rm B}$, but we find the same sensitive switch behavior as before, where a relatively small decrease in $(\eta + \xi)/\alpha_{\rm B}$ can trigger a drastic change in the structure of the disk-corona system, from gas+radiation pressure supported near the equator to overwhelmingly dominated by magnetic support everywhere.  However, the condition for most of the energy flux to escape to $z \gg H$ is now more stringent, $\beta_0 < 1 - \nu$.  
 
Among various simulations extant which show a reversing accretion disk dynamo, we note that the models of \cite{jiang14a} show a much steeper drop in toroidal magnetic pressure, with distance from the disk core, than the $p_{\rm B} \propto z^{-2}$ estimated for our unidirectional field models with $\beta_0 \gg 1$ (see their Fig.~2).  We suggest that reconnection losses across the current sheets separating field reversals are responsible for this steepening.  We note that their space-time diagram showing the ``butterfly" pattern of reversing, buoyant fields (their Fig.~1, bottom panel) shows frequent reversals with one or more large-scale current sheets sweeping through the disk at all times.  Their plots of volume dissipation rate as a function of column density, if fit by an alpha-model with ${\cal H} \propto p_{\rm B}$, allow us to estimate the reconnection efficiency parameter as $\nu \sim 5$ and 8 for models A and B, respectively.  We note, however, that the reconnection in these calculations is driven by numerical resistivity, and does not necessarily reflect the degree of physical reconnection that would actually occur in such a disk.  

This approximate analysis is applicable to cases where field reversals occur on a shorter time scale than the rise time of the flux across a pressure scale height.  Simulations suggest that this is the case for relatively weak toroidal fields that do not provide significant vertical support. For magnetically supported disks that are the main focus of this paper, field reversals appear to be very infrequent \citep[][Salvesen et al., in preparation]{bai13}, and a unidirectional model as developed in Section 2 is appropriate.

\end{document}